\documentclass[sn-mathphys-ay, Namedate, iicol]{sn-jnl}% Math and Physical Sciences Author Year Reference Style
\usepackage{graphicx}    % Including figure files
\usepackage{amsmath}     % Advanced maths commands
\usepackage{amssymb}     % Extra maths symbols
\usepackage{amsfonts}    % Extra maths fonts
\usepackage{xspace}
\usepackage{amsthm}%
\usepackage{mathrsfs}%
\usepackage[title]{appendix}%
\usepackage{xcolor}%
\usepackage{textcomp}%
\usepackage{manyfoot}%
\usepackage{booktabs}%
\usepackage{listings}%
\newcommand{\Msun}{\,$M_{\odot}$\xspace}
\newcommand{\Rsun}{\,$R_{\odot}$\xspace}
\newcommand{\Msyr}{\,$M_{\odot}$\,yr$^{-1}$\xspace}
\newcommand{\kms}{\,km\,s$^{-1}$\xspace}

\newcommand{\cmsqg}{\,cm$^2$\,g$^{-1}$\xspace}

\newcommand{\Ha}{H$\alpha$\xspace}
\newcommand{\Hb}{H$\beta$\xspace}

\newcommand{\FeII}{Fe\,{\sc ii}\xspace}

\newcommand{\A}{\,\AA\xspace}
\newcommand{\apj}{Astrophys. J. } % Astrophysical Journal
\newcommand{\apjl}{Astrophys. J. Lett. } % Astrophysical Journal Letters
\newcommand{\apjs}{Astrophys. J. Suppl. Ser. } % Astrophysical Journal Supplementary Series
\newcommand{\apss}{Astrophys. Space Sci. } % Astrophysics and Space Science
\newcommand{\aap}{Astron. Astrophys. } % Astronomy and Astrophysics
\newcommand{\mnras}{Mon. Not. R. Astron. Soc. } % Monthly Notices of the RAS
\newcommand{\pasp}{Publ. Astron. Soc. Pac. } % Publications of the Astron. Soc. of the Pacific
\begin{document}
%-------------------------------------------------------------------------------
\title[Type IIP supernovae] 
{What stands behind luminosity peak of type IIP supernovae?}
 \author*[1,2]{\fnm{V. P.} \sur{Utrobin}}\email{utrobin@itep.ru}

\author[2]{\fnm{N. N.} \sur{Chugai}}\email{nchugai@inasan.ru}

%\equalcont{These authors contributed equally to this work.}

\affil*[1]{\orgname{NRC ``Kurchatov Institute''},
   \orgaddress{\street{acad. Kurchatov Square 1}, \city{Moscow},
   \postcode{123182}, \country{Russia}}}

\affil[2]{\orgname{Institute of Astronomy, Russian Academy of Sciences},
   \orgaddress{\street{Pyatnitskaya St. 48}, \city{Moscow},
   \postcode{119017}, \country{Russia}}}

\abstract{%
We explore the origin of an initial broad luminosity peak that precedes
   the light curve plateau in type IIP supernovae.
It is well known that the explosion of an evolutionary red supergiant model
   in the frame of radiation hydrodynamics fails to reproduce the initial
   luminosity peak.
A common practice to compensate this drawback is to introduce massive
   ($\gtrsim$0.3\Msun) circumstellar shell attached to the evolutionary model.
This remedy, however, has a negative side effect --- a strong deceleration of
   the outer ejecta, which is inconsistent with early spectral data.
An alternative way out suggests modification of a hydrogen envelope of
   the evolutionary red supergiant model --- the outer layers should have
   steeper density gradient.
We perform the search for the optimal multi-parameter hydrodynamic model of
   type IIP SN~1999em based on the Markov Chain Monte Carlo sampling.
The major result is that the initial luminous peak is well reproduced by
   the explosion of the red supergiant with the logarithmic density gradient
   smaller than $-$3 in the outer layers of about 0.9\Msun.
We suggest that the mechanism for the density steepening in the hydrogen
   envelope is closely related to the unknown mechanism responsible for
   the creation of the ubiquitous low-mass ($<$0.01\Msun) dense confined
   shell detected in type IIP supernovae.
}


\maketitle

%===============================================================================
\section{Introduction}
\label{sec:intro}
%-------------------------------------------------------------------------------
A type IIP supernova (SN~IIP) is an explosion of a red supergiant (RSG) that
   retains a significant part of the hydrogen-rich (hydrogen) envelope until
   the core collapse.
Its light curve is generally understood as the radiative cooling of the
   expanding fireball that ends up about 100 days with the transition to
   the radioactive tail --- the glow of the cold ejecta powered by the
   radioactive decay of $^{56}$Ni $\to ^{56}$Co $\to ^{56}$Fe.
The fireball cooling proceeds through two distinct stages:
   (i) the initial luminosity peak ($t \lesssim 30$ days) followed by
   (ii) the luminosity plateau --- the gradual cooling in the regime of
   a cooling recombination wave (CRW) \citep{GIN_71}.

With a radiation hydrodynamics code in hand (many of them are open sources)
   the light curve and the expansion velocities presumably could be easily
   reproduced by the explosion of an appropriate RSG model.
In fact, the outcome of this naive approach would be disappointing.
Indeed, the explosion of an evolutionary RSG model in the frame of the
   one-dimensional (1D) radiation hydrodynamics lacks the Rayleigh-Taylor
   and turbulent mixing of metals and helium with the hydrogen envelope
   at the (C+O)/He and He/H composition interfaces occurring during the
   explosion.
This mixing dramatically affects the light curve \citep{Woosley_88}.
There is no common recipe for the required mixing in the pre-SN.
Generally, it is advisable to smooth steep density gradients and mix material
   of the different composition in a way that is compatible with the late
   stage of the light curve plateau.

The second failure of the evolutionary RSG model occurs when ones describe
   the initial luminosity peak \citep{UC_08}.  
It is proposed that to reproduce the initial peak requires a significantly
   steeper density gradient in the hydrogen envelope of the RSG model
   than in the evolutionary one \citep{Utrobin_07}.
The mechanism responsible for steepening the density gradient in the RSG
   envelope is unknown.

An alternative solution to the luminosity peak problem is to invoke an
   additional massive ($0.3-1$\Msun) circumstellar (CS) shell attached to
   the evolutionary RSG model \citep{Morozova_2017}.
The last decade this recipe has gained widespread acceptance.
The overlooked drawback of this conjecture is an inevitable deceleration of
   the outer ejecta that turns out inconsistent with the observational data
   \citep[cf.][]{UC_26}.
This flaw rules out the conjecture of a massive CS shell\footnote{%
   Surprisingly, numerous papers promoting the massive CS shell 
   never mention the problem of the ejecta deceleration.}.

The goal of the present communication is to emphasize the point that to solve
   the problem of the early luminosity peak of SNe~IIP one needs to admit that
   the RSG pre-SN has a steeper density gradient of the hydrogen envelope
   compared to the evolutionary RSG model.
As an example, we consider the well-observed SN~1999em ealier studied
   hydrodynamically \citep{Baklanov_05,Utrobin_07}.
The density distribution of the hydrogen envelope will be set by three
   parameters: the inner and outer gradients with the inflection radius
   in between.
This means that with usual four parameters (the ejecta mass, the explosion
   energy,the RSG radius, and the extent of $^{56}$Ni mixing) we obtain
   the seven free parameters for the hydrodynamic model.

To cope with the search for the optimal model in the 7D parameter space,
   we address the Markov Chain Monte Carlo (MCMC) sampling technique.
This procedure requires to perform about $7000-10\,000$ samplings for each run.
It would take 0.5 yr for the radiation hydrodynamics code {\sc CRAB} to perform
   one MCMC run with the available notebook.
We therefore use the 300 times faster code {\sc VELTEM} with the radiation
   transfer in the equilibrium diffusion approximation.

Since the last time the code {\sc VELTEM} was used 30 years ago, we briefly
   reiterate its description (Section~\ref{sec:hydmod}) with some important
   numerical aspects never published before.
In Section~\ref{sec:visfit}, we describe our simple approach to fit the
   hydrodynamic model to the observational data by means of the variation of
   the principal parameters and apply it to SN~1999em.
We then use a more sophisticated approach to get the optimal model
   with the MCMC technique (Section~\ref{sec:mc2mod}).
In Section~\ref{sec:disc} we discuss the obtained results and summarize
   with the conclusion in Section~\ref{sec:concl}.

Below for SN~1999em, we adopt a distance of 11.7 Mpc \citep{LKNT_03},
   a recession velocity of 800 km\,s$^{-1}$ \citep{LFG_02}, a total extinction
   $A_V=0.31$, and an explosion date of JD 2451476.77 \citep{BBH_00, HPM_01,
   EDC_03}.

%===============================================================================
\section{Hydrodynamic model}
\label{sec:hydmod}
%-------------------------------------------------------------------------------
The SN hydrodynamic models discussed below are based on a numerical integration
   of the equations of radiation hydrodynamics in the approximation of
   equilibrium diffusion limit for a gas that moves with spherical symmetry
   in its own gravitational field.
The integration is performed using our radiation hydrodynamics code {\sc VELTEM}
   applied formerly, for example, to SN~1987A \citep{GINU_87, Utr_93}
   and SN~1993J \citep{Utr_94}.

The differential equations of radiation hydrodynamics \citep[e.g.][]{MM_84},
   written for Lagrangian coordinates the time $t$ and the mass $m$ enclosed
   within the sphere of the radius $r$, are:
   the velocity definition,
\begin{equation}
   {\partial r \over \partial t} = v \; ,
\label{eq:eqnvel}
\end{equation}
   the continuity equation,
\begin{equation}
   {\partial r \over \partial m} = {1 \over 4 \pi r^2 \rho} \; ,
\label{eq:eqncont}
\end{equation}
   the momentum equation,
\begin{equation}
   {\partial v \over \partial t} =
   - 4 \pi r^2 {\partial (P + Q) \over \partial m}
   - {G m \over r^2} \; ,
\label{eq:eqnmomt}
\end{equation}
   the energy equation,
\begin{equation}
   {\partial E \over \partial t}
   + (P + Q) {\partial \over \partial t} \biggl( {1 \over \rho} \biggr) =
   - 4 \pi {\partial (r^2 F) \over \partial m} + \varepsilon \; .
\label{eq:eqneng}
\end{equation}
Here $\rho$ is the gas density,
   $P=P_g+E_r/3$ is the total pressure of gas and radiation (the radiative
   internal energy related to the temperature $T$ by $E_r = a T^4$,
   where $a$ is the radiation constant),
   $E=E_g+E_r$ is the total specific internal energy of gas and radiation,
   $Q$ is the artificial viscosity,
   $G$ is the gravitational constant,
   $F$ is the radiative flux, and
   $\varepsilon$ is the specific energy deposition rate, for example, due to
   the radioactive decay of $^{56}$Ni.
To automatically capture shocks, a simple artificial viscosity
   \citep[e.g.][]{RM_67} is added into the hydrodynamic equations.

The diffusion flux of the equilibrium radiation is given by
\begin{equation}
   F_{D} = - 4 \pi r^2 { c \over 3 \chi_{\mbox{\tiny R}}} 
       {\partial E_r \over \partial m} \; ,
\label{eq:fdif}
\end{equation}
   where $c$ is the speed of light and
   $\chi_{\mbox{\tiny R}}$ is the Rosseland mean opacity (true absorption and
   scattering).
The radiative flux $F$, in turn, is represented by the diffusion flux $F_{D}$
   corrected in accordance with a flux-limited diffusion theory of
   \citet{LP_81}:
\begin{equation}
   F = f(\xi) F_{D} \; ,
\label{eq:frad}
\end{equation}
   where $f(\xi)$ is the flux limiter
\begin{equation}
   f(\xi) =  \frac{2 + 3\xi}{2 + 3\xi +3\xi^2} \; , \;
   \xi = \frac{|F_{D}|}{c E_r} \; .
\label{eq:fxi}
\end{equation}
The value $c E_r$ represents the maximum diffusion flux.

As initial conditions for the equations of radiation hydrodynamics we use
   polytropic-like stellar models in hydrostatic equilibrium.
Initial data and boundary conditions are specified to properly formulate
   this problem: the radius $r$, the velocity $v$, and the radiative flux $F$
   at the inner boundary and the radiative flux  $F$ and a zero gas pressure
   ($P_g = 0$) at the outer boundary.

The composition is assumed to be a mixture of hydrogen, helium, and
   an ``average'' heavy element.
At each time step, the ionization equilibrium of this mixture is determined
   by an approximate method of calculating the multiple ionization of
   the heavy element.
The Rosseland mean opacity is evaluated in the hydrogenic approximation,
   allowing for Thomson scattering by free electrons.

Especially for this study we add the contribution of spectral lines to
   the opacity in an expanding medium with a velocity gradient.
It is evaluated by the generalized formula of \citet{CAK_75}.
Oscillator strengths of lines are taken from the line database of
   \citet{Kur_02} containing nearly 530\,000 lines.
Energy level data are from the atomic spectra database of the National
   Institute of Standards and Technology.
Atomic and ionic level populations are determined by the Boltzmann formulae and
   the Saha equations for a mixture of all elements from H to Zn with the local
   temperature.

The transport of $\gamma$-rays from the decay chain
   $^{56}$Ni $\to ^{56}$Co $\to ^{56}$Fe is treated in an approximation
   similar to that of \citet{SW_84} with an effective $\gamma$-ray
   absorption opacity of 0.03 \cmsqg for an electron to nucleon ratio of 0.5.
The latter is inferred by \citet{CPK_80} from Monte Carlo calculations.
Positrons are assumed to deposit their energy locally.

The SN explosion is initiated by a supersonic piston applied to the
   bottom of the stellar envelope at the boundary of the 1.58\Msun
   central core, which is removed from the computational mass domain and
   assumed to collapse to become a neutron star.

To handle the system (\ref{eq:eqnvel}-\ref{eq:eqneng}) described above,
   we replace the differential equations by suitable discrete approximations
   and solve these numerically.
We divide the medium into discrete cells, locating radii, velocities, and
   fluxes at cell interfaces and material and radiation properties
   (density, pressure, energy etc.) at cell centers.
We use an implicit form of both the momentum equation (\ref{eq:eqnmomt})
   and the energy equation (\ref{eq:eqneng}) to integrate the discretized
   equations of radiation hydrodynamics.

It is worth noting that in type IIP SNe during the plateau phase the bolometric
   luminosity is mainly determined by properties of cooling and recombination
   wave (CRW).
This phenomenon was established by \citet{GIN_71} and analyzed in detail by
   \citet{GN_76}.
A main property of the CRW consists in generating virtually the entire energy
   flux carried away by radiation within its front.
This radiation flux exhausts the thermal and recombination energy of
   cooling matter by means of recombination.
It was well demonstrated for SN~1999em \citep{Utrobin_07}.
The numerical treatment of the flux in such specific conditions faces
   difficulties because of the very steep dependence of the opacity on
   the temperature.
The formation of a suitable difference expression for the diffusion flux
   is presented in Appendix~\ref{secA1}.

\begin{table*}[!t]
\caption{Comparison of the optimal models for SN~1999em}
\label{tab:param}
\begin{tabular*}{\textwidth}{@{\extracolsep\fill}lcccccc}
\toprule%
 & & \multicolumn{2}{@{}c@{}}{CRAB} & \multicolumn{1}{@{}c@{}}{VELTEM} & \multicolumn{2}{@{}c@{}}{MCMC} \\\cmidrule{3-4}\cmidrule{5-5}\cmidrule{6-7}%
Parameter & Unit & Value & Error & Value & Value & Error \\
\midrule
Pre-SN radius $R_{pSN}$ & \Rsun & 
500 & $\pm200$ & 600 & 680 & $+46$\,/$-37$ \\
Ejected mass $M_{ej}$ & \Msun &
19.0 & $\pm1.2$  & 22.5 & 21.9 & $+0.7$\,/$-0.7$ \\
Explosion energy $E$ & $10^{51}$\,erg &
1.3 & $\pm0.1$  & 1.61 & 1.45 & $+0.07$\,/$-0.07$ \\
$^{56}$Ni mass $M_{\mathrm{Ni}}$ & \Msun &
0.036 & $\pm0.009$  & 0.036 & 0.036 & -- \\
Extent of $^{56}$Ni mixing $v_{mix}$ & \kms &
 660 & -- & 1360 & 1340 & $+70$\,/$-90$ \\
Inflection point $R_{ip}$ & \Rsun &
 -- & -- & 363 & 378 & $+13$\,/$-13$ \\
Parameter $\alpha$ & -- &
 -- & -- & 0.52 & 0.52 & $+0.02$\,/$-0.02$ \\
Parameter $\beta$ & -- &
 -- & -- & 5.35 & 5.35 & $+0.07$\,/$-0.06$ \\
\botrule
\end{tabular*}
\end{table*}
%
%===============================================================================
\section{Fitting models to data}
\label{sec:twoapp}
%-------------------------------------------------------------------------------
%
%===============================================================================
\subsection{Optimal model via parameter variation}
\label{sec:visfit}
%-------------------------------------------------------------------------------
%
\begin{figure}
   \includegraphics[width=\columnwidth, clip, trim=0 238  51 140]{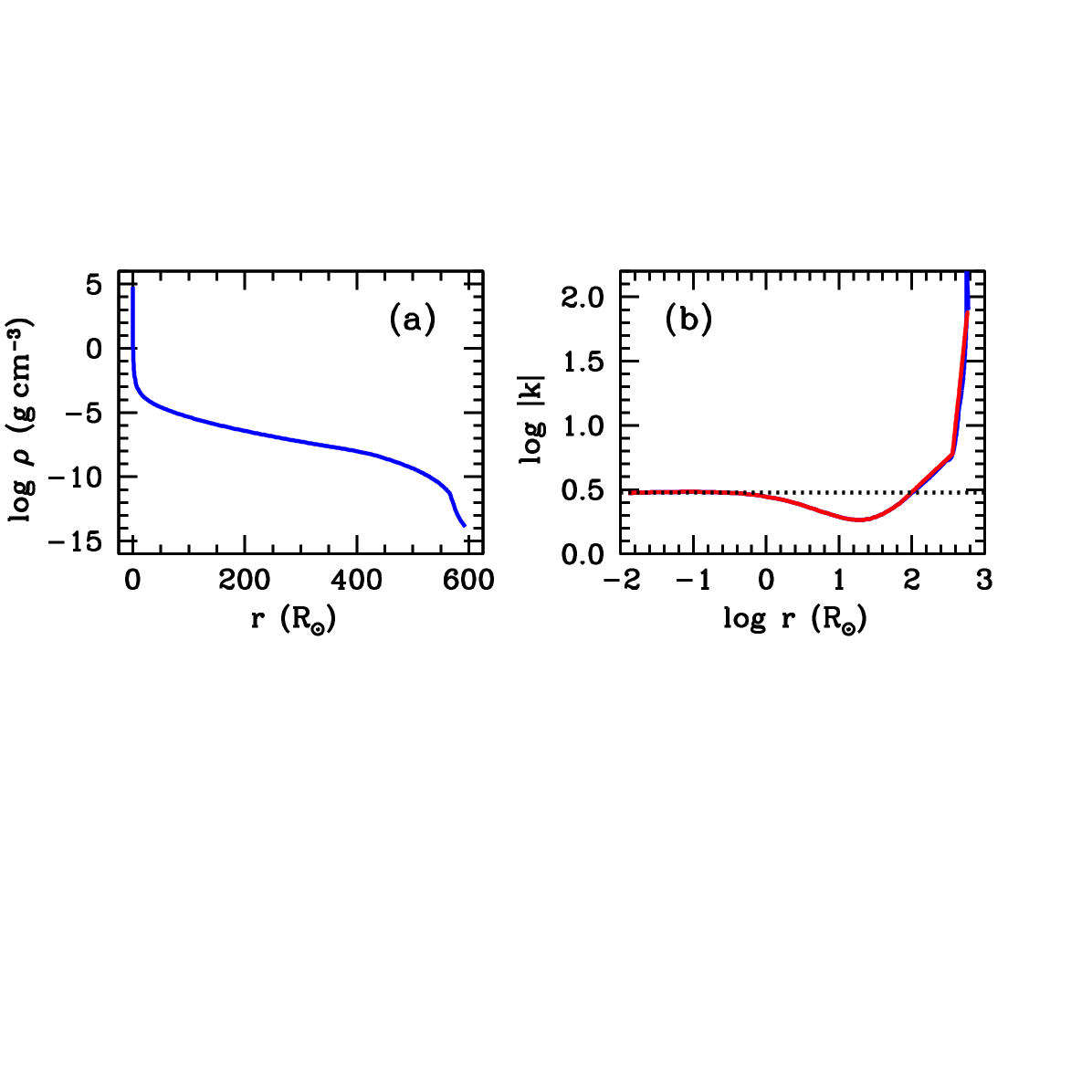}
   \caption{%
   The pre-SN structure of the VELTEM optimal model.
   Panel (a): the density distribution as a function of radius.
   Panel (b): the distribution of the logarithmic density gradient $k$ as
      a function of radius for the same model (\emph{blue line\/}).
    \emph{Red line\/} is a similar pre-SN model the hydrogen
      envelope of which is described by following parameters:
      $R_{base}=62$\Rsun, $R_{ip}=363$\Rsun, $\alpha=0.52$, and
      $\beta=5.35$.
   \emph{Dotted line\/} corresponds to the critical value $k=-3$.
   }
   \label{fig:presn}
\end{figure}
A straightforward search for the optimal model is based on the visual estimate
   of the quality of fitting the observational data.
We used this approach in all our previous hydrodynamic studies of various SNe.
As to SN~1999em, it was earlier investigated in detail with the Lagrangian
   radiation hydrodynamics code {\sc CRAB} \citep{Utrobin_07}.
The inferred basic parameters of the optimal model ---
   the pre-SN radius $R_{pSN}$, the ejecta mass $M_{ej}$, the explosion
   energy $E$, the $^{56}$Ni mass $M_{\mathrm{Ni}}$, and its mixing extension
   $M_{mix}$ --- are listed in Table~\ref{tab:param}.
This optimal model is dubbed the CRAB model.
It is noteworthy that the RSG pre-SN in this case was nonevolutionary with
   a steeper density gradient in the hydrogen envelope compared to that of
   the evolutionary RSG model.

\begin{figure}
   \includegraphics[width=\columnwidth, clip, trim=0 239  54 138]{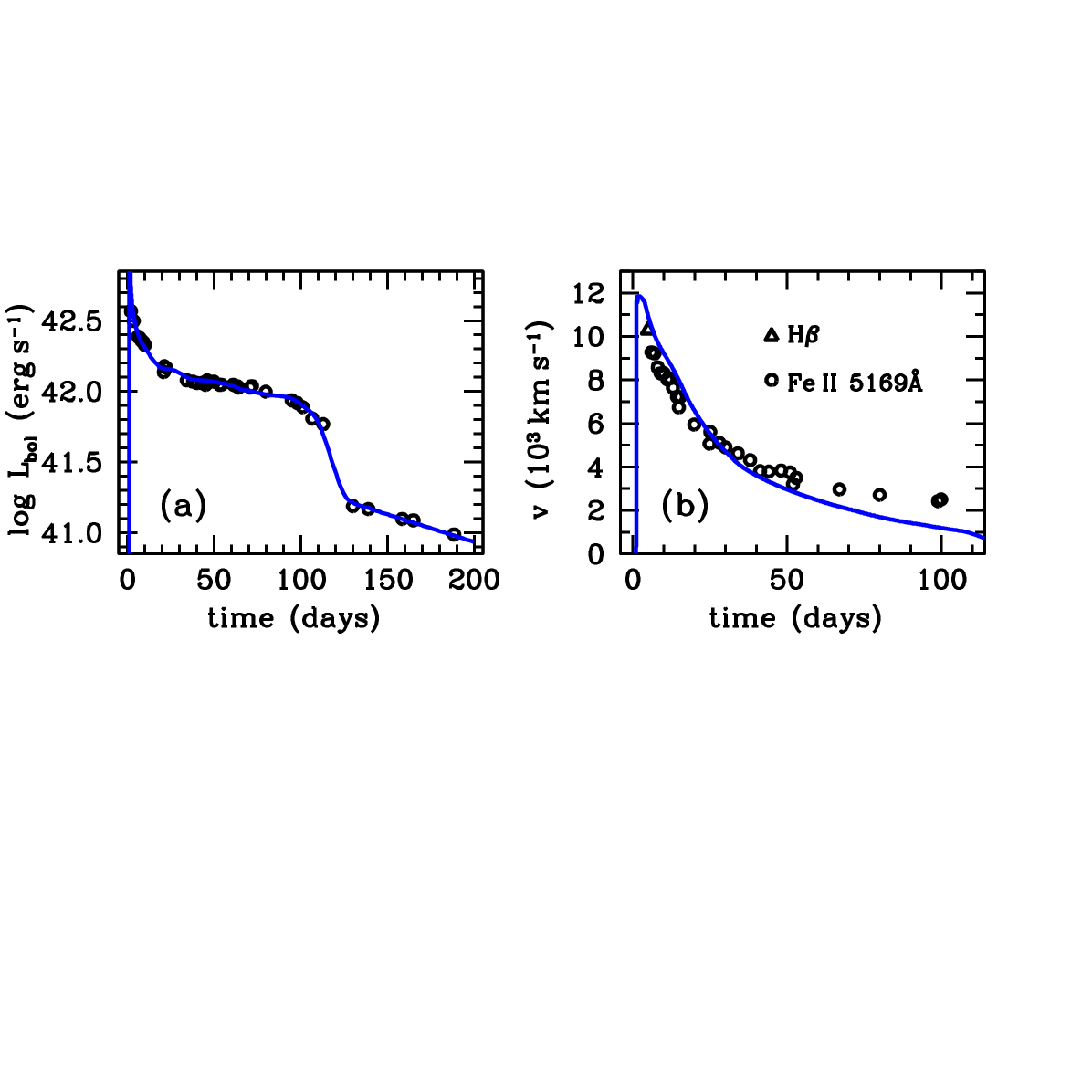}
   \caption{%
   The bolometric light curve and the evolution of photospheric velocity
      for the VELTEM optimal model.
   Panel (a): the model light curve (\emph{blue line\/}) overlaid on
      the bolometric data (\emph{circles\/}) \citep{EDC_03}.
   Panel (b): the evolution of model velocity defined by the level
      $\tau_\mathrm{Thomson} = 1$ (\emph{blue line\/}) is compared with
      the photospheric velocities estimated from the absorption minimum of
      the \Hb and \FeII 5169\A lines \citep{HPM_01, LFG_02}.
   }
   \label{fig:lcv}
\end{figure}
Since here we use a simplified version of the hydrodynamics code {\sc VELTEM}
   with the equilibrium-radiation diffusion (Section~\ref{sec:hydmod}), we
   produce a reference hydrodynamic model of SN~1999em, the VELTEM model.
We construct the pre-SN model (Fig.~\ref{fig:presn}a) similar to that of
   \citet{Utrobin_07}.
The optimal model found via the parameter variation describes the bolometric
   light curve and the expansion velocities (Fig.~\ref{fig:lcv}) for the basic
   parameters listed in Table~\ref{tab:param}.

To explore the effects of the density distribution of the RSG hydrogen envelope
   on the bolometric light curve and the expansion velocities, it is desirable
   to approximate the density profile in a simple way.
Fortunately, the plot $\log |k| - \log r$ (Fig.~\ref{fig:presn}b), where
   $k = d \ln \rho / d \ln r$ is the logarithmic density gradient, reveals
   the two regions with a power law behavior
\begin{equation}
   k(r) = k_0 \left( \frac{r}{r_0} \right)^p \; ,
\label{eq:appr}
\end{equation}
   where $r_0$ and $k_0$ are the radius and the value $k$ at the origin of
   this dependence, respectively, and $p$ is an effective index.
In these inner and outer regions, adjacent at the inflection point of the
   radius $R_{ip}$, the effective index $p$ is given by
\begin{equation}
   p = \left\{
          \begin{array} {ll}
          \alpha & \qquad \textrm{for} \; R_{base}<r<R_{ip} \; , \\
          \beta  & \qquad \textrm{for} \; R_{ip}<r<R_{pSN}  \; , \\
          \end{array}
          \right.
\label{eq:pdef}
\end{equation}
   where $R_{base}$ is the radius close to the bottom of the hydrogen
   envelope.

The above approximation procedure is applied to the density profile of the
   VERTEM model and results in the following parameter values:
   $R_{base}=62$\Rsun, $R_{ip}=363$\Rsun, $\alpha=0.52$, and $\beta=5.35$.
The approximate density distribution (Fig.~\ref{fig:presn}b) agrees well
   with the original distribution that justifies the use of the accepted
   approximation.

It is known that the shock decelerates whenever it encounters a density profile
   that falls off with increasing value $\rho r^3$, while it accelerates for
   density profiles with decreasing value $\rho r^3$ according to
   the blast-wave solution \citep{Sedov_59}.
Dotted line in Fig.~\ref{fig:presn}b indicates the critical value $k=-3$ and
   shows that the velocity of the shock increases when the shock enters the
   layers with the value $k<-3$ within the region $r > 100$\Rsun.
This density distribution guarantees a good fit of the hydrodynamic model to
   the observed light curve and the expansion velocities at the early stage,
   during the first 30 days (Fig.~\ref{fig:lcv}).

%===============================================================================
\subsection{Optimal model via MCMC sampling}
\label{sec:mc2mod}
%-------------------------------------------------------------------------------
%
\begin{figure}[t]
\centering
   \includegraphics[width=\hsize, clip, trim=25  30  27  13]{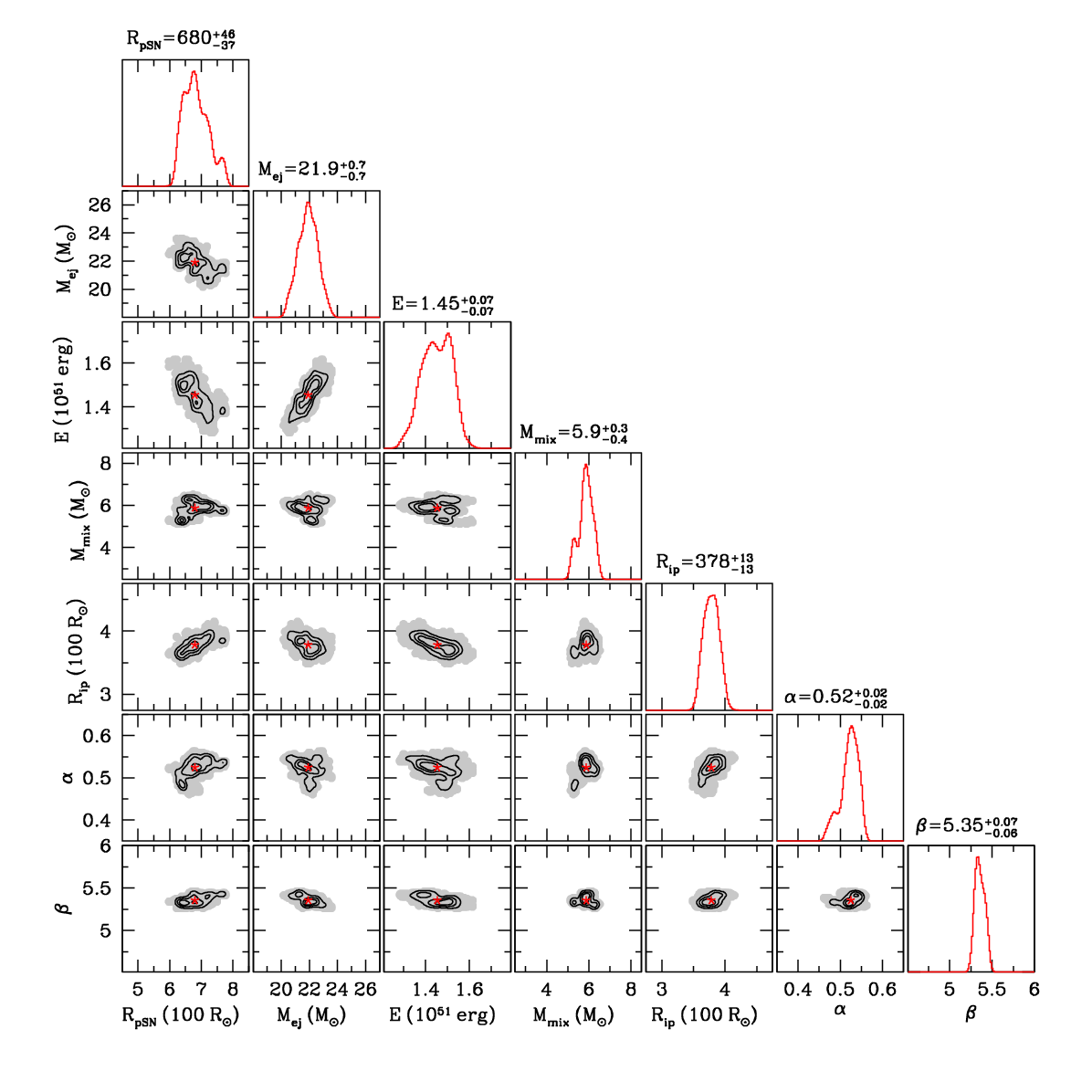}
   \caption{%
   The posterior constraints on the physical parameters of the optimal
      hydrodynamic model for the normal type IIP SN~1999em.
   The contour plots show estimates of the two-dimensional marginalized
      probability densities and the histograms show the marginalized density
      for each parameter.
   The $1\sigma$ confidence intervals centered around the median are given
      at the top of the histograms.
   \emph{Red stars\/} mark the positions of these medians in the two-dimensional
      plots and \emph{gray regions\/} represent the parameter scatter during
      the MCMC sampling.
   }
   \label{fig:cnr_1}
\end{figure}
A more reliable way to optimize the hydrodynamic model is based on the search
   of the likelihood maximum via a random walk in multi-parameter space.
This technique is realized as the MCMC sampling with the Metropolis-Hastings
   acceptance function \citep[e.g.,][]{HF-M_18}.

Our hydrodynamic model for the MCMC study is specified by seven parameters:
   the pre-SN radius $R_{pSN}$, the ejecta mass $M_{ej}$,
   the explosion energy $E$, the $^{56}$Ni mixing extension $M_{mix}$,
   the inflection point radius $R_{ip}$, the power indices $\alpha$ and
   $\beta$.
The first four are the basic parameters in traditional SN hydrodynamic studies,
   whereas other three describe the structure of the hydrogen envelope
   --- a central point of the present study.
The $^{56}$Ni mass $M_{\mathrm{Ni}}$ is not included to the list, because
   its value is fixed and well determined by the luminosity at the radiative
   tail.

As the observational data set, we use the bolometric light curve, the evolution
   of the photospheric velocity, and the maximum velocity of the ejecta.
The ``UBVRI'' bolometric light curve of SN~1999em was constructed by
   \citet{EDC_03}.
The photospheric velocities were evaluated from radial velocities at maximum
   absorption of spectral lines \Hb and \FeII 5169\A measured by
   \citet{HPM_01} and \citet{LFG_02}.
In addition, we estimate the maximum velocity of the SN ejecta of
   $13500\pm500$\kms on JD 2451481.79 from the blue edge of the \Ha absorption
   in the spectrum obtained by \citet{LFG_02}.

The sampling of the parameters is realized as flat priors for all of them.
To evaluate the chi-squared ($\chi^2$) function and, consequently, the
   likelihood function, we compare the calculated hydrodynamic model
   specified by the sampling with the observables of SN~1999em --- the 
   bolometric light curve, the photospheric velocities, and the maximum
   velocity of the ejecta (Fig.~\ref{fig:lcv}).

We start our sampler with the seven parameters of the VELTEM model
   in Table~\ref{tab:param} and then the MCMC code makes 7000 steps.
This results in posterior probability distribution functions (pdfs) for these
   parameters and two-dimensional scatter plots for each parameter pairs shown
   in the corresponding corner plot (Fig.~\ref{fig:cnr_1}).
The medians of these posterior pdfs and the $1\sigma$ confidence intervals
   centered around the medians are given in the corner plot and
   Table~\ref{tab:param}.
Note that the two-dimensional scatter plots $R_{pSN}-M_{ej}$, $R_{pSN}-E$,
   and $M_{ej}-E$ show correlation between parameters which is in a agreement
   with the earlier study \citep{Utrobin_07}.

\begin{figure}
   \includegraphics[width=\columnwidth, clip, trim=0 239  54 138]{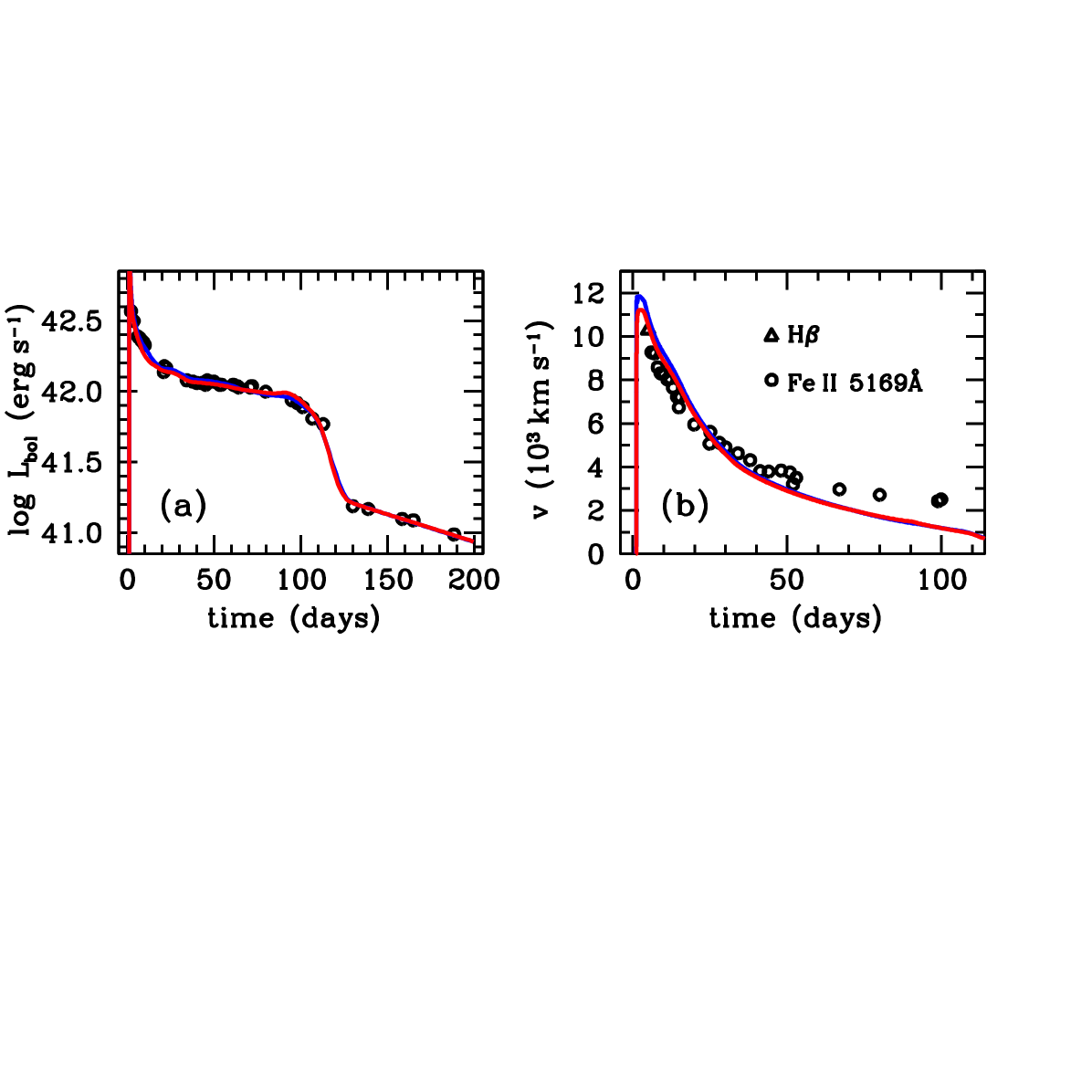}
   \caption{%
   A comparison of the MCMC optimal model (\emph{red line\/}) with
      the VELTEM optimal model (\emph{blue line\/}).
   See Fig.~\ref{fig:lcv} legend for details.
   }
   \label{fig:clcv}
\end{figure}
The inferred optimal parameters (Table~\ref{tab:param}) are used as the input
   data for the hydrodynamic code {\sc VELTEM} to produce the light curve and
   the photospheric velocities and then verify them with the observations
   (Fig.~\ref{fig:clcv}).
It turns out that the results obtained with the MCMC posterior pdfs are 
   indistinguishable from those achieved by the visual optimization
   (Fig.~\ref{fig:lcv}).
This is convincing demonstration that (i) the medians of the posterior pdfs,
   in fact, represent the MCMC optimal model of SN~1999em and (ii) the visual
   optimization used previously in all hydrodynamic studies with the code
   {\sc CRAB} is a reliable procedure.

Moreover, the optimal model found by the MCMC sampling recovers the density
   distribution of the hydrogen envelope that is responsible for the
   description of the initial luminosity peak (Fig.~\ref{fig:clcv}).
Note that the characteristics of this density profile are very close to those
   of the VELTEM optimal model (Table~\ref{tab:param}).
This supports the key idea that the steep density gradient of the hydrogen
   envelope is a mandatory prerequisite for the RSG presupernova of SNe~IIP.

\begin{figure}[t]
\centering
   \includegraphics[width=\hsize, clip, trim=25  30  27  13]{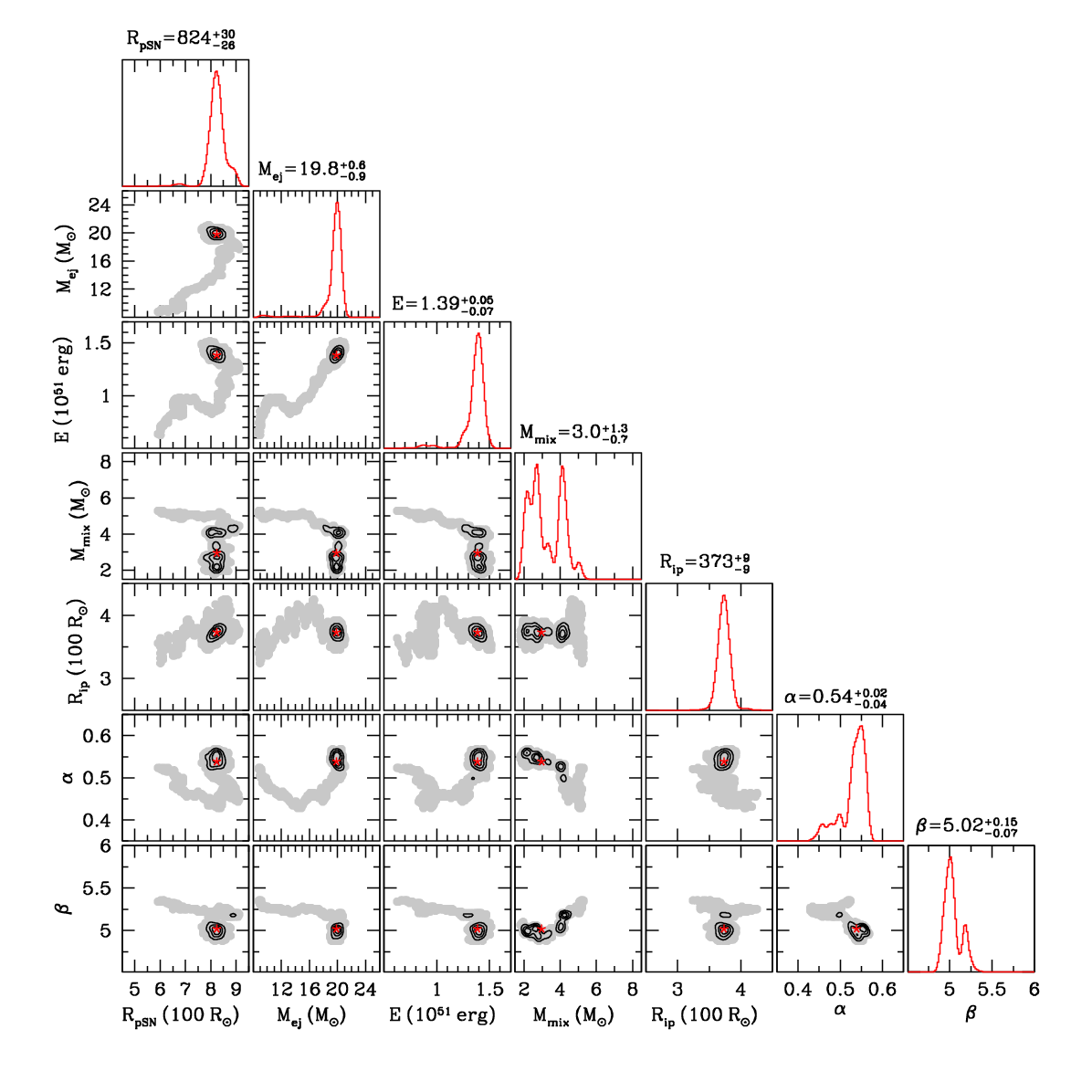}
   \caption{%
   The travel of the SN hydrodynamic model from a low-mass configuration to
      the global extreme in the seven parameters space.
   See Fig.~\ref{fig:cnr_1} legend for details.
   }
   \label{fig:cnr_2}
\end{figure}
Bearing in mind that the outcome of random walk in multi-parameter space might
   depend on the starting point, it is advisable to check, whether we could
   miss the solution with a low-mass progenitor.
To investigate this problem, we initialize the sampling with the same set of
   parameters as in the previous MCMC run but a low ejecta mass $M_{ej}=9$\Msun
   and a low explosion energy $E=0.6\times10^{51}$\,erg.
After 10\,000 steps, these two parameters eventually relax at
   the values of a high ejecta mass and a high explosion energy comparable
   to those of the MCMC optimal model (Fig.~\ref{fig:cnr_2}).
The corner plot demonstrates the absence of hints of a low-mass pre-SN.
We conclude, therefore, that the radiation hydrodynamics modeling rules out
   the pre-SN of SN~1999em significantly less massive than the earlier
   estimate of about 21\Msun \citep{Utrobin_07}.

%===============================================================================
\section{Discussion}
\label{sec:disc}
%-------------------------------------------------------------------------------
%
\begin{figure}
   \includegraphics[width=\columnwidth, clip, trim=15 113  25 116]{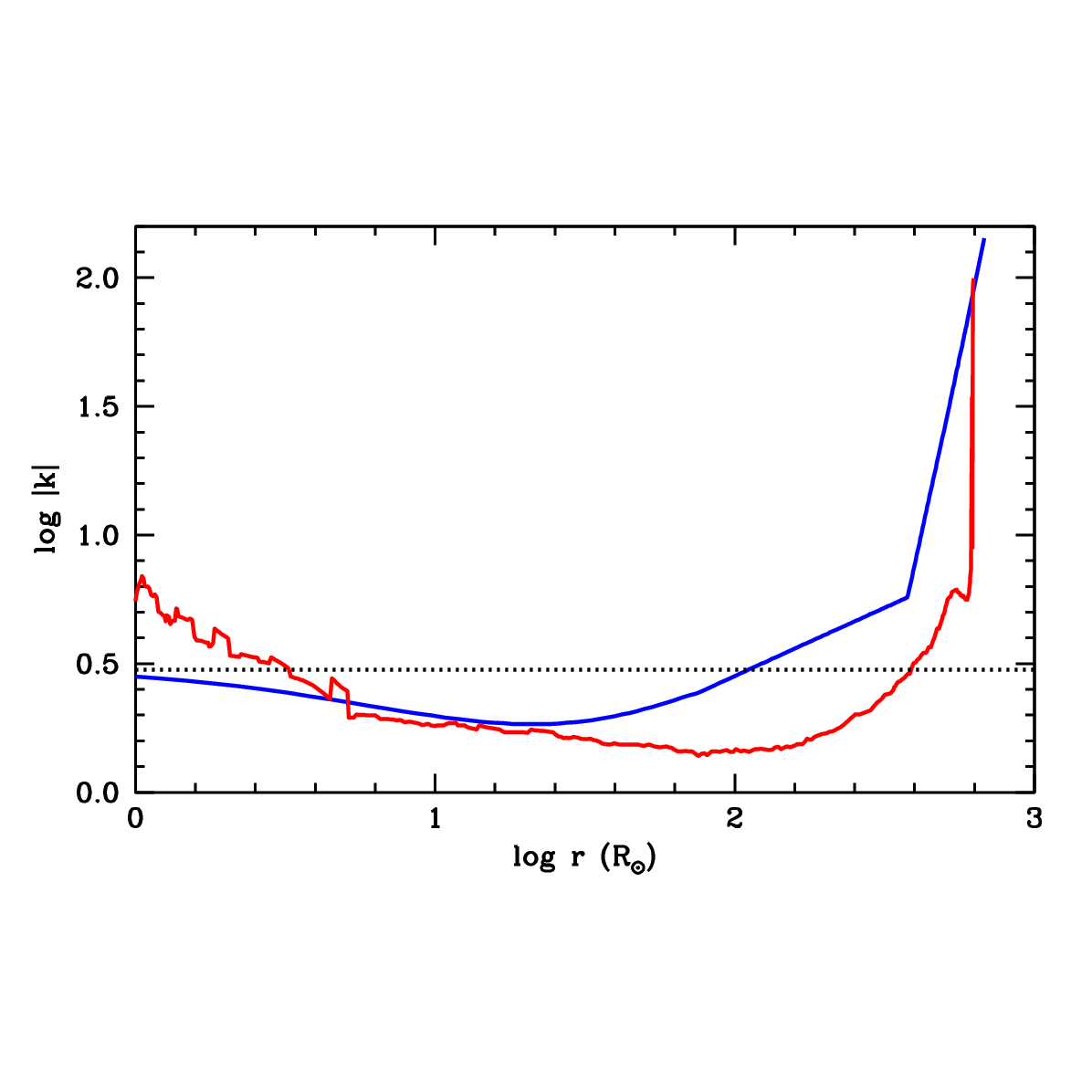}
   \caption{%
   The distribution of the logarithmic density gradient $k$ as a function of
      radius for the MCMC pre-SN model (\emph{blue line\/}).
    \emph{Red line\/} is the similar distribution for the evolutionary
      pre-SN L15 model of \citet{LSC_00}.
   \emph{Dotted line\/} corresponds to the critical value
      $k= d \ln \rho / d \ln r = -3$.
   }
   \label{fig:evolmod}
\end{figure}
The optimal model of the well-observed type IIP SN~1999em produced by
   the MCMC sampling of posterior pdfs for all seven parameters supports
   the conjecture that the steep density distribution of the RSG hydrogen
   envelope is responsible for the initial luminosity peak.

The comparison between the density distribution of the hydrogen envelope in
   the nonevolutionary pre-SN for the MCMC optimal model and that of
   the evolutionary RSG L15 model \citep{LSC_00} is shown in terms of
   the gradient $k(r)$ (Fig.~\ref{fig:evolmod}).
The plot demonstrates that in the range of $100-600$\Rsun the evolutionary RSG
   has the gradient $k > -3$ which indicates that the shock wave in this
   region decelerates, whereas the MCMC optimal model requires $k < -3$ in
   the same range of radii.
The steep density gradient in the outer layers of RSG favors the luminous early
   light curve via producing the conditions that increase a fraction of
   the escaped radiation compared to the radiation adiabatic losses
   in the expanding ejecta.
Note that the mass of the external ($r > 100$\Rsun) layers with the gradient
   $k < -3$ in the MCMC pre-SN model is only 0.88\Msun.
Hence, the density distribution of the hydrogen envelope in the evolutionary
   RSG model should be modified in the outer layers of about 0.9\Msun
   to fit the luminosity peak observed in SNe~IIP.

A mechanism for the formation of a steep density gradient of the RSG hydrogen 
   envelope remains to be understood. 
Yet the spectroscopic data for SN~II at the early ($t < 1$ week) stage
   indicate hydrodynamic processes that take place shortly before
   the SN explosion.
Indeed, there is a strong evidence for an enhanced mass loss during the last
   years ($\lesssim10$) before the core collapse.
This phenomenon is manifested by the emission lines of hydrogen, helium,
   and other species with specific ``narrow core + broad wings'' that
   originate in the CS dense confined ($r < 10^{15}$ cm) shell (DCS) with
   the Thomson optical depth of $\sim$3 \citep{Chugai_2001, Groh_2014,
   Yaron_2017}. 
The estimated DCS mass lies in range of $0.003-0.007$\Msun \citep{Yaron_2017,
   UC_26} which suggests the tremendous mass-loss rate $\sim$10$^{-3}$\Msyr
   during $5-10$ years before the SN explosion.
We suggest that the mechanism responsible for the formation of the low-mass DCS
   could also produce steepening the density gradient in the outer layers of
   the RSG hydrogen envelope.

Earlier \citep{UC_08} we faced a clear disparity between the cumulative
   distribution of SNe~IIP progenitors in the $9-25$\Msun mass range
   derived from the Salpeter IMF and the distribution of hydrodynamic
   progenitor masses of SNe~IIP.
This disparity revealed a pronounced lack of SNe~IIP progenitors with
   the mass lower than 15\Msun.
The growing number of well-observed SNe~IIP studied hydrodynamically
   in a uniform approach does not improve this situation.
The current view on this troubling issue was given by \citet{UC_25}.
Given this fundamental problem, of particular interest is the question of
   an existence of the SN~1999em pre-SN less massive than that of the MCMC
   optimal model.
We answer this question with the special MCMC run confirming that our radiation
   hydrodynamics studies exclude the pre-SNe of SN~1999em significantly less
   massive than about 21\Msun \citep{Utrobin_07}.

For many years in our hydrodynamic studies we used the visual optimization
   and the term ``the optimal model'' to name our best model, admitting
   that it is not optimal model in mathematical sense.
This study based on the MCMC method convinces us that the visual optimization
   may be as good as the MCMC sampling, if we take into account
   all important observables: the bolometric light curve, the evolution of
   the photospheric velocity, and the maximum velocity of the ejecta.

%===============================================================================
\section{Conclusion}
\label{sec:concl}
%-------------------------------------------------------------------------------
We have performed the search for the optimal model of the type IIP SN~1999em
   based on the MCMC sampling posterior pdfs for seven parameters including
   three parameters describing structure of the RSG hydrogen envelope.
The major result is that the initial broad luminous peak is well reproduced
   by the explosion of the RSG with a steeper density gradient of the hydrogen
   envelope compared to that of the evolutionary RSG model.

\backmatter

\bmhead{Acknowledgements}
The work of V.U. was carried out within the framework of the state
   assignment of the National Research Center ``Kurchatov Institute''.

\bmhead{Author contribution}
The authors contributed equally to this work.

\bmhead{Funding}
Not applicable.

\bmhead{Data availability}
No datasets were generated or analyzed during the current study.

\bmhead{Materials availability}
Not applicable.

\bmhead{Code availability} 
Not applicable.

\section*{Declarations}

\subsection*{Ethics approval and consent to participate}
Not applicable.

\subsection*{Consent for publication}
Not applicable.

\subsection*{Competing interests}
The authors declare no competing interests.

%===============================================================================
\begin{appendices}
%-------------------------------------------------------------------------------
\section{ }
\label{secA1}
%-------------------------------------------------------------------------------
Here we construct an adequate difference expression for the diffusion flux
   (\ref{eq:fdif}), developing the original ideas of \citet{Chet_85}.

To begin with, we represent the difference diffusion flux $S$ at
   the interface $j$ between the mass cells $\Delta m_{j-1/2}$ and
   $\Delta m_{j+1/2}$ with the radiative internal energies
   $U_{j-1/2}$ and $U_{j+1/2}$, respectively, in the following simple form:
\begin{equation}
   S_j^{smp} = - { 8 \pi c \over 3 } r_{j}^2 { U_{j+1/2}-U_{j-1/2} \over 
         \Delta m_{j-1/2}\chi_{j-1/2}+\Delta m_{j+1/2}\chi_{j+1/2} } \; ,
\label{eq:ssmp}
\end{equation}
   which was obtained by \citet{Marchuk_77}.
We rewrite this expression as
\begin{equation}
   S_j^{tau} = - { c \over 2\sqrt{3} } { U_{j+1/2}-U_{j-1/2} \over 
         \Delta \tau_{j-1/2}+\Delta \tau_{j+1/2} } \; ,
\label{eq:stau}
\end{equation}
   introducing the optical depths of the corresponding cells
\begin{subequations}\label{eq:dtau}
\begin{align}
   \Delta \tau_{j-1/2} & = { \sqrt{3} \over 16 \pi r_{j}^2} \Delta m_{j-1/2} 
                         \chi_{j-1/2} \\
   \Delta \tau_{j+1/2} & = { \sqrt{3} \over 16 \pi r_{j}^2} \Delta m_{j+1/2} 
                         \chi_{j+1/2} \; .
\end{align}
\end{subequations}%\par

As mentioned above, the CRW involves the recombination front which separates
   the optically thick medium from the thin one.
It is clear that at the CRW position the expression (\ref{eq:stau}) gives
   unacceptable result when $\Delta \tau_{j-1/2} \to \infty$ and
   $\Delta \tau_{j+1/2} \to 0$.
To overcome this problem, \citet{Chet_85} solved the radiative diffusion
   equation at the vicinity of two adjacent cells and obtained the correct
   difference expression for the diffusion flux.
We apply this approach to spherically symmetric geometry and represent
   a general form of the difference expression which is applicable both
   at the location of a steep change in the optical depth and outside it.

The radiative diffusion equation \citep[e.g.][]{ZR_67} is represented as
\begin{subequations}\label{eq:eqdiff}
\begin{align}
   { dL \over dm } + c \chi (U-U^*)  & = 0 \\
   16 \pi^2 r^4 { c \over 3 \chi } { dU \over dm } + L & = 0 \; ,
\end{align}
\end{subequations}
   where $L=4 \pi r^2 S$ is the luminosity  and $U^*=aT^4$ is the equilibrium
   radiative internal energy.
The difference solution of this equation for the mass interval [$m_j$,$m_{j+1}$]
   with the boundary conditions $U=U_j$ and $L=4 \pi r_j^2 S_j$ at $m=m_j$
   and $U=U_{j+1}$ and $L=4 \pi r_{j+1}^2 S_{j+1}$ at $m=m_{j+1}$ is
\begin{subequations}\label{eq:soltn}
\begin{align}
    \widetilde{S}_{j+1}-\widetilde{S}_j + c (U_{j+1}+U_j-2 U_{j+1/2}^*) A_{j+1/2} & = 0 \\
   c (U_{j+1}-U_j) + (\widetilde{S}_j + \widetilde{S}_{j+1}) A_{j+1/2} & = 0 \; ,
\end{align}
\end{subequations}
   where
\begin{eqnarray}
   \widetilde{S}_j = { \sqrt{3} \over 4 \pi r_{j+1/2}^2} L_j  \; , \qquad
   \widetilde{S}_{j+1} = { \sqrt{3} \over 4 \pi r_{j+1/2}^2} L_{j+1}  \; , \nonumber \\
   A_{j+1/2} = \tanh{\left({ \sqrt{3} \over 8 \pi r_{j+1/2}^2} \Delta m_{j+1/2} 
                         \chi_{j+1/2}\right)} \nonumber \; .
\label{sedge}
\end{eqnarray}
Applying the equation (\ref{eq:soltn}b) to the mass intervals [$m_{j-1/2}$,$m_j$]
   and [$m_j$,$m_{j+1/2}$] and assuming sensibly $L_j = L_{j-1/2} = L_{j+1/2}$
   and $r_j = r_{j-1/4} = r_{j+1/4}$, we obtain the correct difference expression
   for the diffusion flux at the interface $j$ between the optically thick and
   optically thin cells
\begin{equation}
   S_j^{new} = - { c \over 2\sqrt{3} } { U_{j+1/2}-U_{j-1/2} \over 
         \tanh{\Delta \tau_{j-1/2}}+\tanh{\Delta \tau_{j+1/2}} } \; .
\label{eq:snew}
\end{equation}
In contrast to the equation (\ref{eq:stau}), for the equilibrium radiation
   this expression results in physically sensible limit
   $S_j^{new} \to (2/\sqrt{3}) \sigma T_{j-1/2}^4$ ($\sigma = ac/4$)
   when $\Delta \tau_{j-1/2} \to \infty$ and $\Delta \tau_{j+1/2} \to 0$.

For the case when both $\Delta \tau_{j-1/2} \to \infty$ and
   $\Delta \tau_{j+1/2} \to \infty$, however, the obtained expression
   (\ref{eq:snew}) is unsatisfactory.
To construct a difference expression which is valid both at the location of
   a steep change in the optical depth and outside it, we define the weight
   factor
\begin{equation}
   q =  P(\Delta \tau_{j-1/2}) P(\Delta \tau_{j+1/2}) \; ,
\label{eq:q}
\end{equation}
   where the auxiliary function
\begin{equation}
   P(z) = { \tanh{z}-z+z^2 \over 1-z+z^2} \;
\label{eq:pz}
\end{equation}
   has the following properties:
\begin{equation}
%\begin{eqnarray}
   P(z) \approx \left\{
          \begin{array} {ll}
          2z^5/15                        & \textrm{for} \; z \ll 1 \; , \\
          1-{2\exp(-2z) \over 1-z+z^3/3} & \textrm{for} \; z  >  1 \; . \\
          \end{array}
          \right.
\label{eq:pzlim}
%\end{eqnarray}
\end{equation}
The corresponding properties of the weight factor $q$ permit us to specify
   the effective optical depth
\begin{eqnarray}
   \Delta \tau_j^{eff} = q (\Delta \tau_{j-1/2}+\Delta \tau_{j+1/2}) + \nonumber\\
   (1-q)(\tanh{\Delta \tau_{j-1/2}}+\tanh{\Delta \tau_{j+1/2}})
\label{eq:taueff}
\end{eqnarray}
   and, consequently, a general expression of the diffusion flux
\begin{equation}
   S_j^{gen} = - { c \over 2\sqrt{3} } { U_{j+1/2}-U_{j-1/2} \over 
                 \Delta \tau_j^{eff}  } \; .
\label{eq:sgen}
\end{equation}
\end{appendices}

%===============================================================================

%% if required, the content of .bbl file can be included here once bbl is generated
%%\input sn-article.bbl
%-------------------------------------------------------------------------------
\end{document}